\documentclass[11pt]{article}
\usepackage{setspace}
\usepackage{amssymb}
\usepackage{amsmath}
\usepackage{amsfonts}
\usepackage{slashed}

\def\be{\begin{equation}}

\def\ee{\end{equation}}

\def\bea{\begin{eqnarray}}

\def\eea{\end{eqnarray}}

\makeatletter
\def\blfootnote{\xdef\@thefnmark{}\@footnotetext}
\makeatother

\begin{document}

\singlespace

\begin{flushright} BRX TH-6639 \\
CALT-TH 2018-038
\end{flushright}

\vspace*{.3in}

\begin{center}

{\Large\bf A Pedagogical Note on Matter Covariantization for Gravity Coupling}

{\large S.\ Deser}

{\it 
Walter Burke Institute for Theoretical Physics, \\
California Institute of Technology, Pasadena, CA 91125; \\
Physics Department,  Brandeis University, Waltham, MA 02454 \\
{\tt deser@brandeis.edu}
}
\end{center}

\begin{abstract}
Covariantization is of course required for initially flat space matter actions to couple consistently to GR; here we show in detail for concrete systems how it follows in the same physical way as that deriving GR from its initial free field form.
\end{abstract}

Pure GR can be physically (rather the geometrically) derived as a unique self interacting system from an initially free spin $2$ gauge field [1] in arbitrary backgrounds [2], a mandatory step if matter sources are to be included. The parallel, also necessary, ``covariantizing" of the sources was proved there by the following short argument: Since the matter stress tensors are simultaneously defined as the variations of their actions with respect to their (say flat) background geometry $\eta^{\mu\nu}$, and their variations with respect to the spin $2$ field $h^{\mu\nu}$ to which they couple, this functional relation immediately requires their actions to depend only on their sum, the background-independent metric $g= \eta+h$. It is perhaps useful to  amplify the physics of our procedure in terms that apply to any interacting matter-field-system. The stress-tensor $T_{\mu\nu}$ is simply a current --- i.e.\, the matter action $A$'s response to a change in the (background) gravitational field, $T = \delta A/\delta \eta$; this current $T$ must then be coupled to $h$ via $hT$ and added to $A$ to provide that interaction. The process must be repeated as long as any lone $\eta$ remains, because each yields a further current contribution. Clearly, the result is the universal replacement $A=\int L(\hbox{matter}; \eta, h)\rightarrow  A=\int L (\hbox{matter}; g=\eta+h)$ where $\eta$ and $h$ have the same, built-in, tensorial transformation properties. This is precisely the principle of how one obtains the coupling of matter to gauge fields, an initial fictitious $A_\mu$ potential replacing $\eta$. The above proof was not explicit enough for some, hence this pedagogical note, with the (rather pedestrian) details for massive or massless fields of spins $(0,1)$. Higher spin matter is omitted, as it has other problems coupling to gravity. We also omit spinors, only because they would take too much space; the self-coupling derivation of SUGRA [3] took care of spin $3/2$. The process is entirely similar to that for GR, with the major technical exception that an infinite number of steps (or the above functional argument) are needed for most models because they contain ``naked" $\sqrt{-\eta} \equiv \sqrt{-\det{\eta}}$.

Perhaps surprisingly, matter is best treated here (and most like pure GR in first order form) in second order form, where the respective actions read
\begin{eqnarray}
I[\phi] &=& -\frac{1}{2} \int dx \left[ D_\mu \phi^* D_\nu \phi \eta^{\mu\nu} + m^2 \phi^* \phi \sqrt{-\eta}\right], \\
I[A] &=& -\frac{1}{4} \int dx \biggl[ F_{\mu\nu} F_{\alpha\beta} \left( \eta^{\mu\alpha} \right) \left( \eta^{\nu\beta} \right)/\sqrt{-\eta} +2 m^2 A_\mu A_\nu \eta^{\mu\nu} \biggr]. 
\end{eqnarray}
Here $F_{\mu\nu}(A)$ is the usual YM or Maxwell field strength, and we have allowed for the scalar's eventual gauge coupling through a covariant vector gauge (not metric) derivative $D_\mu$.  The background dependence is through the same contra-density combination as in the GR construction, except for a extra $1/\sqrt{-\eta}$ in the $F^2$ term and a $\sqrt{-\eta}$ in the scalar's mass term; note that our $\det{\eta}$'s determinant is just that of the covariant $\eta$ tensor. We use Rosenfeld's definition of a system's stress tensor as a generalized current --- the variation of its action with respect to the $\eta$ as if they were Riemann metrics of appropriate weight, before restoring them to their original status. [Any other definition would not alter the results, as shown in [1].] Thus the covariantization is a one-step process in the scalar kinetic and vector mass term, where adding the coupling $h^{\mu\nu} T_{\mu\nu}$ leaves the metric dependence precisely in the background-independent combination $g^{\mu\nu} =\eta^{\mu\nu} + h^{\mu\nu}$, while it takes just one more step to covariantize the $F^2$ term since there are two $\eta$ pieces there. The $\sqrt{-\eta}$ terms in $F^2$ and in the scalar mass can most expeditiously be dispatched using the functional variation equalities discussed at the outset, or more pedantically by showing the same thing term by term in an infinite series expansion of $\sqrt{-\eta}$, adjoining the required factor of $h$ at each step to reach $\sqrt{-g}$. Pure GR did not have this extra problem, depending only on $h^{\mu\nu}$. Note the (unsurprising, in view of scalar-tensor mixing) exception of massless scalars, which are one-step, like GR.

Were we to treat matter, as we did gravity, in first order form instead --- the gravitational field's simplest incarnation --- things would not simplify. The flat space spin $0$ and $1$ actions are now respectively
\begin{eqnarray}
I[\phi] &=& -\int dx \left[ \pi^{*\, \mu} D_{\mu} \phi - \frac{1}{2} \pi^{*\, \mu} \pi^\nu \eta_{\mu\nu} + \frac{1}{2} m^2 \phi^* \phi \sqrt{-\eta}\right], \\
I[A] &=& -\frac{1}{2} \int dx \biggl[ {\mathcal F}^{\mu\nu} F_{\mu\nu}(A) -\frac{1}{2} {\mathcal F}^{\mu\nu} {\mathcal F}^{\alpha\beta} \eta_{\mu\alpha} \eta_{\nu\beta} \sqrt{-\eta} 
+ m^2 A_\mu A_\nu \eta^{\mu\nu} \biggr].
\end{eqnarray}
These fields' actions would not involve gravitational covariant derivatives in any background. Here, $\pi^\mu$ and ${\mathcal F}^{\alpha\beta}$ are independent variables.  Note that the kinetic terms are metric-independent, so do not contribute to the respective (again Rosenfeld)  stress-tensors. We must assign density weights to the respective variables; again the results are choice-independent, but it is most convenient to make the field momenta contravariant densities and the amplitudes covariant scalars and vectors respectively. Then in (3), the $\pi^2$ term's metric, $\eta_{\mu\nu}$  is the inverse of a contra-density; similarly, in (4), the ${\mathcal F}^2$ term picks up a $\sqrt{-\eta}$ in addition to two inverse metric densities.  The resulting stress tensors are the standard ones for these models; most important, their initial coupling to gravity --- replacing their $\eta$ by the $h^{\mu\nu}$ (densities), i.e., adding $h^{\mu\nu} T_{\mu\nu}$ to the actions automatically yields the correct final coupling (only) to the full contravariant metric density $g^{\mu\nu} = \eta^{\mu\nu} +h^{\mu\nu}$  (or its inverse of course), as is trivially verified for both spins: thus in (3), the first step in gravity coupling adds $\pi \pi H$, where $H$ is the component for the field as a covariant anti-density, not to be confused with the inverse of $h$, whose existence is not assumed! So the process ends there in terms of the inverse contravariant full metric $\eta+H$ (recall that the transformation nature of the $h$-field component to be used is determined by that of the original $\eta$, to maintain coordinate invariance); there are two independent covariantizations proceeding simultaneously: in $h$ and in $H$, whose relations are as infinite series. The explicit $\sqrt{-\eta}$'s may be dealt with as in the second order discussion. In (4), the ${\mathcal F}^2$ term has two, inverse contra density $\eta$, which requires two steps to fully covariantize, first to get $\eta (\eta+H)$ then again to promote this to $(\eta+H)^2$.  We have thus shown that reaching correct matter-gravity coupling is uniform with reaching gravity self-coupling, albeit a bit more complicated!

\subsubsection*{Acknowledgements}
This work was supported by grant DOE\#desc0011632; I thank J. Franklin for major tech help.

\end{document}